\documentclass{kluwer}    
\usepackage{epsfig}

\newcommand\apj{ApJ}

\newcommand\apjl{ApJ (Lett)}

\newcommand\mnras{MNRAS}

\newcommand{\ba}{$(b/a)^2_{\rm bar}$}

\begin{document}
\begin{article}
\begin{opening}

  \title{Quantifying Morphological Evolution from Low to High Redshifts}
  \author{Roberto G. \surname{Abraham}}
  \runningauthor{R. G. Abraham}
  \runningtitle{Quantitative Morphology}
  \institute{Institute of Astronomy, University of Cambridge,
    Madingley Road, Cambridge, CB3 OHA, United Kingdom}

  \begin{abstract} Establishing the morphological history of ordinary
    galaxies was one of the original goals for the {\em Hubble Space
    Telescope}, and remarkable progress toward achieving this this
    goal has been made. How much of this progress has been at the
    expense of the Hubble sequence? As we probe further out in
    redshift space, it seems time to re-examine the underlying
    significance of Hubble's tuning fork in light of the the
    spectacular and often bizarre morphological characteristics of
    high redshift galaxies.  The aim of this review is to build a
    morphological bridge between high-redshift and low-redshift galaxy
    populations, by using quantitative morphological measures to
    determine the maximum redshift for which the Hubble sequence
    provides a meaningful description of the galaxy population. I will
    outline the various techniques used to quantify high-redshift
    galaxy morphology, highlight the aspects of the Hubble sequence
    being probed by these techniques, and indicate what is getting
    left behind. I will argue that at higher redshifts new techniques
    (and new ideas) that place less emphasis on classical morphology
    and more emphasis on the link between morphology and resolved
    stellar populations are needed in order to probe the evolutionary
    history of high-redshift galaxies.  \end{abstract}


\end{opening}

\section{Introduction}

In an influential series of lectures delivered nearly two decades ago,
Kormendy (1982) described how morphological considerations supply the
basic framework for our understanding of galaxies, but also noted that

\begin{quote}
  ``.... morphology is more generally a `soft' science, which is best
  viewed as preparation for more quantitative work. Its most important
  use may be to provide a list of specific questions which provide
  direction for this work.''
\end{quote}

In the same series of lectures, Kormendy espoused the view that {\em
physical morphology} (a scheme in which morphological components such
as lenses, bars, disks and halos are superposed in order to build up a
coherent classification for galaxies) provides a promising avenue
toward the ultimate goal of a taxonomy of galaxies that is physically
interpretable.

In this review I will try to take Kormendy's ideas a little further
and argue that with the advent of deep imaging data from the {\em
Hubble Space Telescope} (HST), coupled with advances in objective
galaxy classification and measurement, the subjective art of
morphological classification has begun to yield to the quantitative
science of physical morphology.  The new perspectives offered by
observing distant galaxies {\em in situ} allows physical morphology to
go straight to the heart of issues central to our understanding of
galaxy evolution.  In this review I will focus on several of these key
issues, posed in the form of the following three questions: (1) At
what redshift is the Hubble sequence observed to be in place?  (2)
Does the Hubble sequence contain its own ``ground state'', or do
entirely new classes of galaxy emerge at early look-back times?  (3)
What drives morphological evolution (ie.  which components of galaxies
form first)?  Data that offers insight into the first of these
questions now exists, and will be reviewed in \S2. The second question
is presently wide open, so in \S3 I will describe the measurement of
promising physical parameters that should at least help constrain the
possibilities. In \S4 I will focus on the third question, and describe
the role of resolved stellar populations in reconstructing the physics
of galaxy formation at high redshifts.

\section{At What Redshift is the Hubble Sequence in Place?}

\subsection{What are we really measuring?}

The impressive physical correlations along the Hubble sequence are
discussed extensively in several recent reviews (Sandage \&
Bedke~1994; Roberts \& Haynes~1994; van~den~Bergh~1998;
Abraham~1999d). The physical significance of the Hubble sequence is
a testament to the genius of Hubble, and to the hard work of
subsequent generations of astronomers who have elaborated and
refined Hubble's original system  over the past 75 years (eg.
Sandage 1961, 1981; de Vaucouleurs et al. 1976, 1991; van den
Bergh 1960, 1976). But before going on to consider recent evidence
for evolution of the Hubble sequence as a function of redshift, it
is first necessary to shed some light on a fact that will shock
nobody working with HST data, but might come as a surprise to
those focusing on morphology in the low-redshift Universe. {\em
The Hubble sequence is a wholly unsuitable basis for the study of
high-redshift galaxies}. As a result of this, most workers
analyzing deep HST data have adopted private galaxy classification
schemes (such as the ten-bin MDS system, or the ubiquitous
three-bin early/late/peculiar system adopted by many groups) that
attempt to preserve the spirit of Hubble's tuning fork (Hubble
1926, 1936), but ride roughshod over the details. This seems to me
to be a reasonable strategy, for three reasons:

\begin{enumerate}

\item{\bf The Hubble sequence is not robust at low signal-to-noise
levels.}  Because the bulges of galaxies generally correspond to
regions of high surface brightness, one of the three central
parameters of the Hubble sequence (bulge-to-disk ratio) is far
more robust than the other two parameters defining the system
(pitch angle and resolution of spiral arms). As a result,
classifications of high-redshift galaxies are generally based on
apparent bulge-to-disk ratio, with little (and often no) regard to
the visibility of spiral arms. This is true regardless of the
detailed mechanics (visual inspection, profile fitting, neural
nets, decision trees, etc) of the classification process.  Studies
of spiral structure at high redshifts are in their infancy (see
\S2.3). In fact, it can be argued that classifications of high
redshift galaxies have more in common with Morgan's Yerkes system
(Morgan 1958, 1959), based on central concentration of light, than
with Hubble's original system.

\item{\bf Some archetypal high redshift galaxies do not exist in the
Hubble sequence.}  Recent work from deep HST imaging surveys has
shown that much of the faint galaxy population is comprised of a
diverse set of morphologically peculiar galaxies entirely outside
the Hubble system. These may be tidally distorted counterparts to
conventional tuning fork galaxies, or entirely new classes of
objects. It seems to me deeply unsatisfactory to simply bin
one-third of the galaxy population (at $I=25$ mag) together into a
catch-all ``peculiar'' category, but that is the current state of
the art. Another concern is that Hubble sequence is intended to
describe only  luminous galaxies, which is appropriate for bright
local surveys where most detected systems are near $L_\star$.
However, deeper magnitude-limited surveys {\em may} (depending on
the faint-end slope of the luminosity function) probe much farther
down the luminosity function so that typical systems are
substantially fainter than $L_\star$. Therefore some component of
a perceived evaporation of the tuning fork towards higher redshift
could be alternatively interpreted as the evaporation of our local
$L_\star$ window function.

\item{\bf Even for local galaxies, the consistency between visual
morphological classifications is poor.}  The results from
controlled comparisons between independent morphological
classifications of local galaxies made by expert morphologists
\cite{Naim:1995} are depressingly bad.  The upshot of this is that
one must have rather serious doubts about observer-to-observer
consistency in visual classifications at {\em any} redshift even
though there is little doubt in my mind that an individual expert
morphologist can make classifications that are internally highly
consistent. When combined with the complications (quantified
below) introduced by a variable rest-wavelength of observation for
high-redshift galaxies, these factors force me to conclude that
objective machine-based classifications are essential for progress
in this field.

\end{enumerate}

In light of these considerations, when I review in the next section
the results from studies intended to probe evolution along the Hubble
sequence, it is important for the reader to bear in mind that what is
really being probed is evolution in the early--late axis of the tuning
fork, indirectly traced by bulge-to-disk ratio. In the local Universe
this parameter is correlated with, but does {\em not} define, position
along the tuning fork.

\subsection{The early--late axis of the tuning fork}

\begin{figure}
\begin{center}
  \epsfig{figure=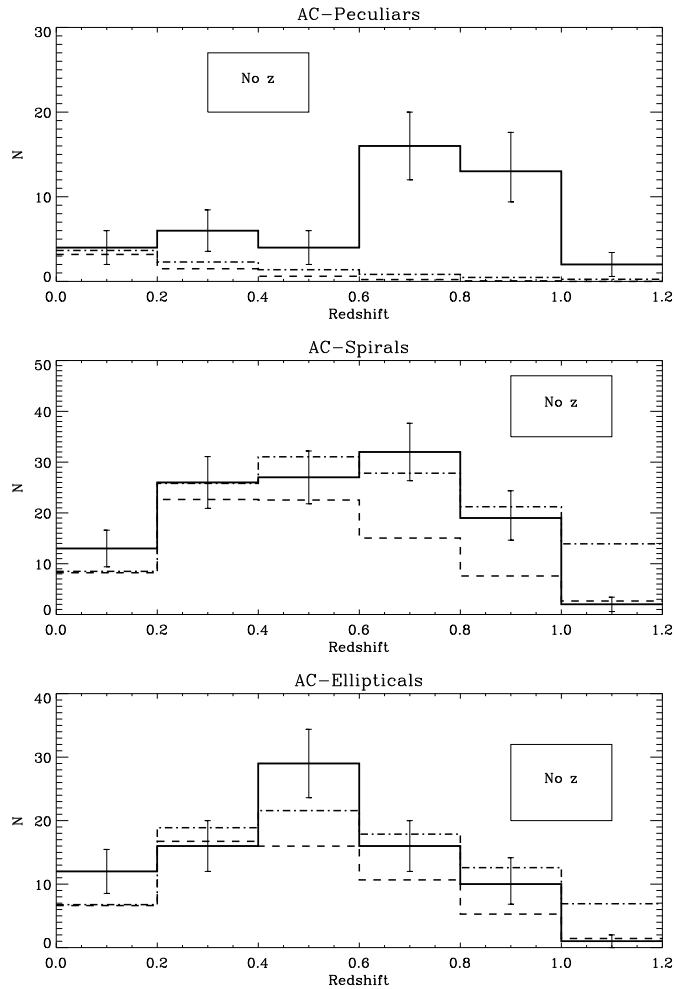,width=9.5cm}
  \caption[NZ]{Morphologically segregated number counts from Brinchmann
    {\em et al.} (1998) \cite{Brinchmann:1998}, based on data from the
    CFRS/LDSS collaboration. The solid-line bins show counts as a
    function of redshift for irregular/peculiar/merger systems (top),
    spirals (middle), and ellipticals (bottom). Morphological
    classifications have been made from WF/PC2 images using an automated
    technique based on central concentration and asymmetry of galaxian
    light \cite{Abraham:1996a}.  The model curves have been
    corrected for ``morphological K-corrections'',
    accounting for the effects of observing the galaxies at bluer rest
    wavelengths as a function of redshift. Superposed on the observed
    histograms are the predictions of no-evolution (dashed) and 1 mag
    linear evolution to $z=1$ (dot-dashed) models. At $z\sim 1$
    approximately 40\% of the galaxy population is morphologically
    peculiar.}
  \label{fig:nz}
\end{center}
\end{figure}

The clearest evidence for morphological evolution to $z \sim 1$ does
not come from work on the Hubble Deep Field, but from the deep HST
imaging follow-ups to the CFRS and LDSS redshift surveys (Brinchmann
et al. 1998; Lilly et al. 1998). The statistical completeness of the
underlying redshift surveys (Lilly et al.  1996; Ellis et al. 1996)
are very well understood, and at $z<1$ cosmological bandshifting
serves mostly to synchronize observed $I$-band data closer to rest
wavelength $B$-band, where galaxy morphology is most familiar. The
morphologically segregated (on the basis of machine
classifications\footnote{Recent years have seen great advances in the
usefulness of objective, machine-based morphological classifications
designed to probe position along the early--late axis of the tuning
fork \cite{Doi:1993, Naim:1995, Abraham:1996a, Odewahn:1996}. Even the
comparatively simple two-parameter system with which I am most
familiar is able to classify galaxies with an accuracy comparable to
that achieved by the visual classifications of independent observers
(Abraham et al. 1996b; Brinchmann 1998).})  $I_{AB} < 22.5$ mag
number--redshift distributions from Brinchmann {\em et al.} (1998) are
shown in Figure~1, along with the predictions of simple no-evolution
and mild-evolution models. These authors conclude that there is little
evidence for substantial evolution in the early-type population, some
evidence for modest evolution in the spiral population (consistent
with mild luminosity evolution at 0.5--1 mag level by $z=1$), and a
spectacular over-abundance of systems categorized as peculiar. These
results seem quite convincing, although the apparently good agreement
with the models should not be over-interpreted, as the absolute
normalization of the local luminosity function is probably known only
to within a factor of two.

These results are in good agreement with earlier findings obtained by
other groups studying the morphologically segregated number-magnitude
relations from the Medium Deep Survey
\cite{Glazebrook:1995,Driver:1995,Abraham:1996a}. The redshift
distributions lay to rest earlier concerns that the luminous peculiar
population at $I<23$ mag is severely contaminated by ``morphologically
K-corrected'' very high-redshift systems, and allows calculation of
the size function for subsets of the data \cite{Lilly:1998}.  It is
not clear whether these trends are consistent with models based on
hierarchical growth (although at a basic level the results do appear
consistent with the semi-analytical prescription of Baugh, Cole \&
Frenk 1996).  If there is a peak of star formation activity at
$z\simeq$1-2 (Madau 1998), either the evolutionary behaviour of
massive regular galaxies beyond $z\simeq$1 must change dramatically
from that observed for $z<1$, or perhaps the trends delineated from
the optical photometric data are underestimated because of
complications such as dust extinction (Meurer et al 1997).
Intriguingly, there seems to be no change in the space density of
large spiral systems to redshift $z=1$, in apparent contradiction of
the predictions of hierarchical models, although attempts are being
made to reconcile the sizes of ``big disks'' with theory
\cite{Mao:1998}.

Beyond $I\sim 22-23$ mag, the HDF is required for reliable
morphological classifications. Various authors (Abraham et al
1996a; Mobasher et al 1996; Odewahn et al 1996; Driver et al 1998)
have used the HDF to extend the earlier Medium Deep Survey
morphological source counts to fainter limits, finding that the
fraction of the peculiar systems increases to at least
$I_{814}=$25 mag, and finding some evidence for a turn-over in the
early-type counts beyond $I=24$ mag (Abraham et al 1996a). Another
striking result from the HDF is the small angular size of the
faintest sources, implying physical extents of only 2-4 $h^{-1}$
kpc.  In a recent analysis, Bouwens et al (1998) suggest that such
sizes cannot be reconciled with the expected redshift and surface
brightness dimming of typical $z<$1 sources and thus claim
substantial physical growth and merging must have occurred for the
galaxy population during the redshift range 1$<z<$3. In another
recent paper, Driver et al.  (1998) have used photometric
redshifts to extend the CFRS/LDSS number-redshift distributions to
$I=26$ mag. These authors confirm both the earlier claims of a
deficit of early-type systems beyond $I=24$ mag, and the rapid
rise in the proportion of morphologically peculiar systems with
redshift. Driver et al. (1998) also note that beyond $z=2$ very
few well-ordered spirals are seen (see the review by Windhorst in
these proceedings), in agreement with the basic point made by van
den Bergh et al. (1996). Beyond $z=1$ a comparison with physical models
becomes problematic. In addition to the uncertain normalization of
the local luminosity function, at least three other factors come
into play: (1) Evolutionary $K$-corrections become very uncertain,
especially for sub-$L_\star$ systems; (2) The appropriate
cosmology is unknown (eg. the possible presence of
$\Omega_\Lambda$); (3) The importance of dust is unknown.

On the basis of all these results, it seems to me that one can
safely place the epoch at which the early--late axis of the Hubble
sequence is well-established to be somewhere around $z \sim 1$.
At this redshift most luminous spirals and ellipticals are in
place, but at least 30\% of luminous galaxies lie off the
sequence, and the sequence as a whole cannot really be considered
a reasonably complete description of the luminous galaxy
population\footnote{Note that I am excluding low surface
brightness systems from consideration at this stage, although
constraints on the putative existence of these in the HDF are now
very severe, as discussed in \S3}.

\subsection{The tines of the tuning fork}

I will now consider the redshift at which the ``orthogonal'' axis of
the tuning fork, namely the bifurcation into barred and unbarred
systems, is established.  As described earlier, studies of spiral
structure at high redshifts are in their infancy. While simple
quantitative measures of overall galactic structure are adequate for
placing even faint HDF galaxies within a one-dimensional early--late
classification sequence, much higher signal-to-noise data are needed
to probe spiral structure.  Another fundamental complication is that
the fraction of {\em local} barred spirals is poorly established;
there is clearly a continuum in apparent bar strength, and the
strength required to merit classification as a barred galaxy is highly
subjective. These difficulties can be side-stepped by using
quantitative measures of bar strength on appropriate subsamples of
very high signal-to-noise data spanning a broad range in redshift
space, so that no reference to visually determined estimates of the
local bar fraction is needed.

\begin{figure}
  \centerline{\epsfig{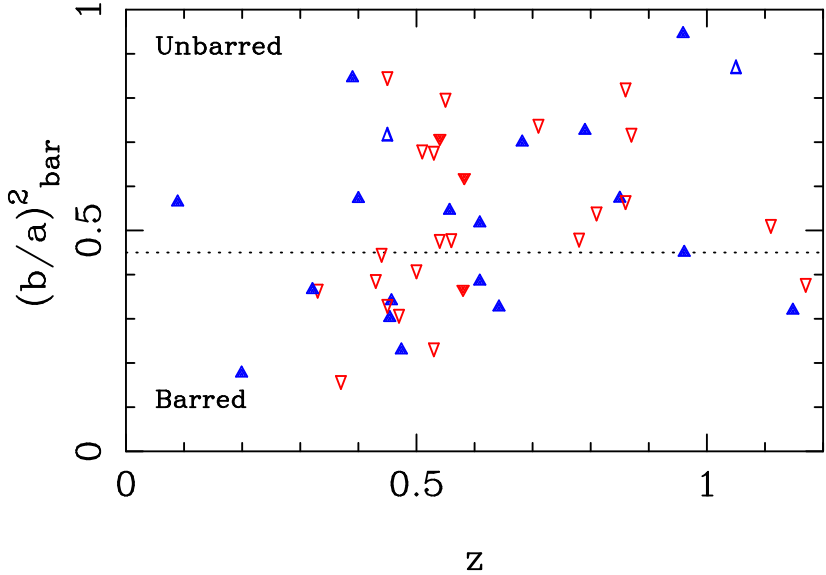}}
  \caption{The bar strength estimator $(b/a)^2_{\rm bar}$ plotted as
    a function of spectroscopic redshift (filled symbols) or
    photometric redshift (open symbols) in the northern (triangles
    pointing upward) and southern (triangles pointing downward) Hubble
    Deep Fields. Note the marked gradient in the proportion of barred
    systems with redshift, beginning at $z\sim 0.5-0.6$. Taken from
    Abraham et al. (1999b).}
\end{figure}

Abraham et al. (1999b) combine data from the Northern and Southern
HDF fields in order to define a sample of luminous $I<23.2$ mag
low-inclination spirals with redshifts less than $z \sim 1$, where
bandshifting effects are negligible and where signal-to-noise
levels are high enough for unambiguous bar detection. Bar strength
is quantified using a simple measure, \ba, corresponding to the
{\em physical} axial ratio of a bar under the assumption of an
elliptical bar embedded within a round, thin disk.  Where these
assumptions are not a good approximation, the estimator still
yields a perfectly quantitative, objective parameter that appears
to closely track visual estimates of bar strength.  For reasonably
low-inclination systems ($< 60$ degrees), spirals classed locally
as SA in the sample of Frei et al. (1996) are cleanly-separated
from systems classed as SB.  The results obtained by applying this
statistic to the HDF data are shown in Figure~2.

This figure reveals a striking decline in the proportion of barred
examples beyond a redshift $z \sim 0.5$. This cannot be due to
uncertainties in using photometric redshifts for the Southern HDF
data, as the same effect is seen in both HDF samples and at
$I<23.2$ mag the Northern HDF is spectroscopically complete at the
90\% level. Similarly, this result cannot be due to bandshifting
effects: the redshift range considered corresponds to the
rest-frame optical, and the effect is most striking at redshifts
close to rest-frame $B$-band, where the barred spiral fraction is
best established in the local Universe. This is confirmed from an
analysis of the northern HDF, where deep NICMOS observations allow
the construction of a sample imaged at a uniform rest-frame
(Abraham et al. 2000).  Formally, the redshift distributions of
the barred and unbarred samples selected on the basis of \ba~ in
Figure~2 are inconsistent at the 99.8\% confidence level from a
Kolmogorov-Smirnov test.

Figure~2 would seem to bring forward the epoch at which the
conventional Hubble system is observed to be in place, from $z
\sim 1$ (based the early--late axis of the tuning fork, as
described in the previous subsection) to $z \sim 0.5$. The
physical mechanisms responsible for the absence of barred spirals
at high redshifts is unclear.  Obvious possibilities include
dynamically hotter (or increasingly dark-matter dominated)
high-redshift disks, and an enhanced efficiency in bar destruction
at high redshifts (Sellwood 1999; Combes 1999).

\subsection{Summary}

Hubble's tuning fork appears to provide a good description of the
morphologies of luminous galaxies out to redshifts $z \sim 0.5$, at
which point the proportion of strongly barred spirals begins to drop
off. By a redshift $z=1$ around 30\% of galaxies within two magnitudes
of $M_\star$ lie off the early--late axis of the classification
sequence.

\section{Alternative Metrics of Galaxy Morphology}

Early results from deep NICMOS observations of the Hubble Deep Field
(Thompson et al. 1998; Bunker et al. 1999) have lain to rest the
possibility that most high-redshift morphologically peculiar systems
are simply late-type spiral galaxies whose peculiar appearance is due
to their being imaged in the rest-frame ultraviolet. The bizarre
appearance of most of these galaxies is intrinsic, and classification
schemes which encompass the diversity in the forms observed are needed
to realistically capture the appearance of galaxies in the distant
Universe.  However, an appropriate classification scheme for such
systems should not just be descriptive.  The ultimate goal of galaxy
classification is to mirror an underlying physical order in the
systems being studied, and alternative approaches to galaxy
classification that are appropriate for studying distant galaxies
should therefore take into account not only the practical limitations
on resolution and signal-to-noise, but also our improved understanding
of the physical basis for galaxy formation.

I would argue that central concentration (or bulge-to-disk ratio) is a
useful (albeit crude) probe of the relative importance of thermally
supported versus rotationally supported structures, and can be used to
test rather directly the predictions of hierarchical formation models.
Measures of bar strength capture (again, rather crudely) information
regarding the dynamical state of the disk, and probe the importance of
secular processes in building up galaxies.  Similarly, on the basis of
a close correspondence to underlying physics, I would argue that bulk
asymmetry and mean surface brightness should be considered fundamental
morphological characteristics of high-redshift galaxies.

\subsection{Asymmetry}

Locally, most morphologically peculiar systems show dynamical
evidence for tidal disruption, and it is tempting to assume that a
large fraction of the diverse peculiar galaxy population seen on
deep images are actually mergers in progress. Few of the
high-redshift peculiar systems in the Hubble Deep Field resemble
the classical appearance of local merging systems, but the usual
signatures of mergers (eg.  tidal tails) are no longer visible at
$z>2$, and at these redshifts the effects of bandshifting on
morphology can be rather extreme. Merging starburst systems seem
to provide at least qualitatively reasonable counterparts to many
faint peculiar galaxies (Hibbard \& Vacca 1997). On the other
hand, synchronization in the colours of some morphologically
peculiar systems (eg. in the ``chains'' first identified by Cowie,
Hu, \& Songaila 1995) seems difficult to explain as the product of
mergers (Abraham 1999c), and many morphologically peculiar systems
do not resemble archetypal starbursts on deep NICMOS data probing
rest-frame optical light (Bunker et al. 1999).  Therefore a
crucial issue in high-redshift studies is to distinguish between
intrinsically asymmetric protogalactic systems and mergers in
progress, and perhaps at a higher level to consider whether such a
distinction is even meaningful in the context of hierarchical
formation scenarios. I don't think I can describe the situation
more clearly than I did in my Les Houches lectures:

\begin{quote}
  ``When should an amorphous blob of components be regarded as single
  morphologically peculiar galaxy, as opposed to a system of
  interacting proto-galaxies?  de~Vaucouleurs used to dismiss the
  notion of considering mergers to be fundamental morphological units
  with the observation that `car wrecks are not cars'. But when the
  road is littered with wrecks, and when the by-product of a wreck is
  another working car, it may be time to re-assess the wisdom of
  restricting morphological classification to regular-looking
  systems''.
\end{quote}

Studies of asymmetry at high redshifts may shed light on this
issue. Conselice et al. (1997) and Takamiya et al. (1999a,b)
find strong correlations between color and symmetry in both local
and high-redshift galaxies. Intriguingly, Conselice et al. (1999)
also demonstrate that at least some interacting systems can be
distinguished from other classes of morphologogically peculiar
systems on the basis of position on a color-asymmetry plane. Using
imaging data from the LDSS/CFRS survey, Le F\`evre et al. (1997; 1999)
show how objective measures of bimodality can parameterize the
growth in the merger rate with redshift, finding evidence for a
sharp increase in the merger rate with redshift. It is worth
noting however that this study was restricted to relatively low
redshifts ($z<1$) and, like all such studies to date, may not
account fully for the evolution in the background counts needed in
order to translate observed close pair counts to a physical
merging fraction (R. Carlberg, private communication).  I suspect
that the ultimate clarification of the nature of the
morphologically peculiar systems in general, and the most distant
Lyman-break systems (Steidel et al. 1997; Giavalisco, Steidel, \&
Macchetto 1996) in particular, must await the completion of at
least the first round of dynamical studies on these objects. These
studies are needed to provide the basic physical framework upon
which future morphological work can be built.

\subsection{Mean Surface brightness}

\begin{figure}
  \centerline{
    \epsfig{file=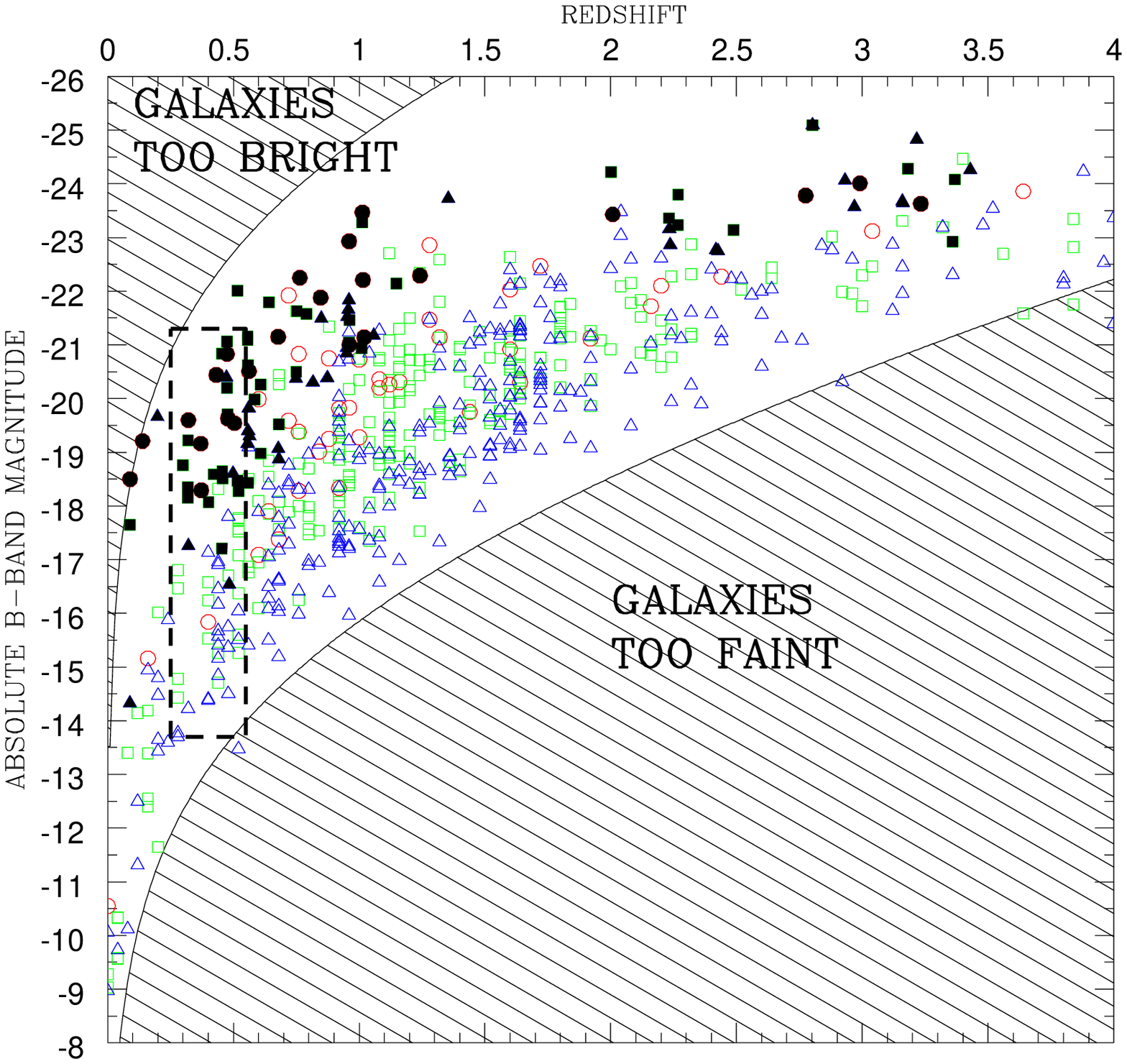,width=16pc}
    \epsfig{file=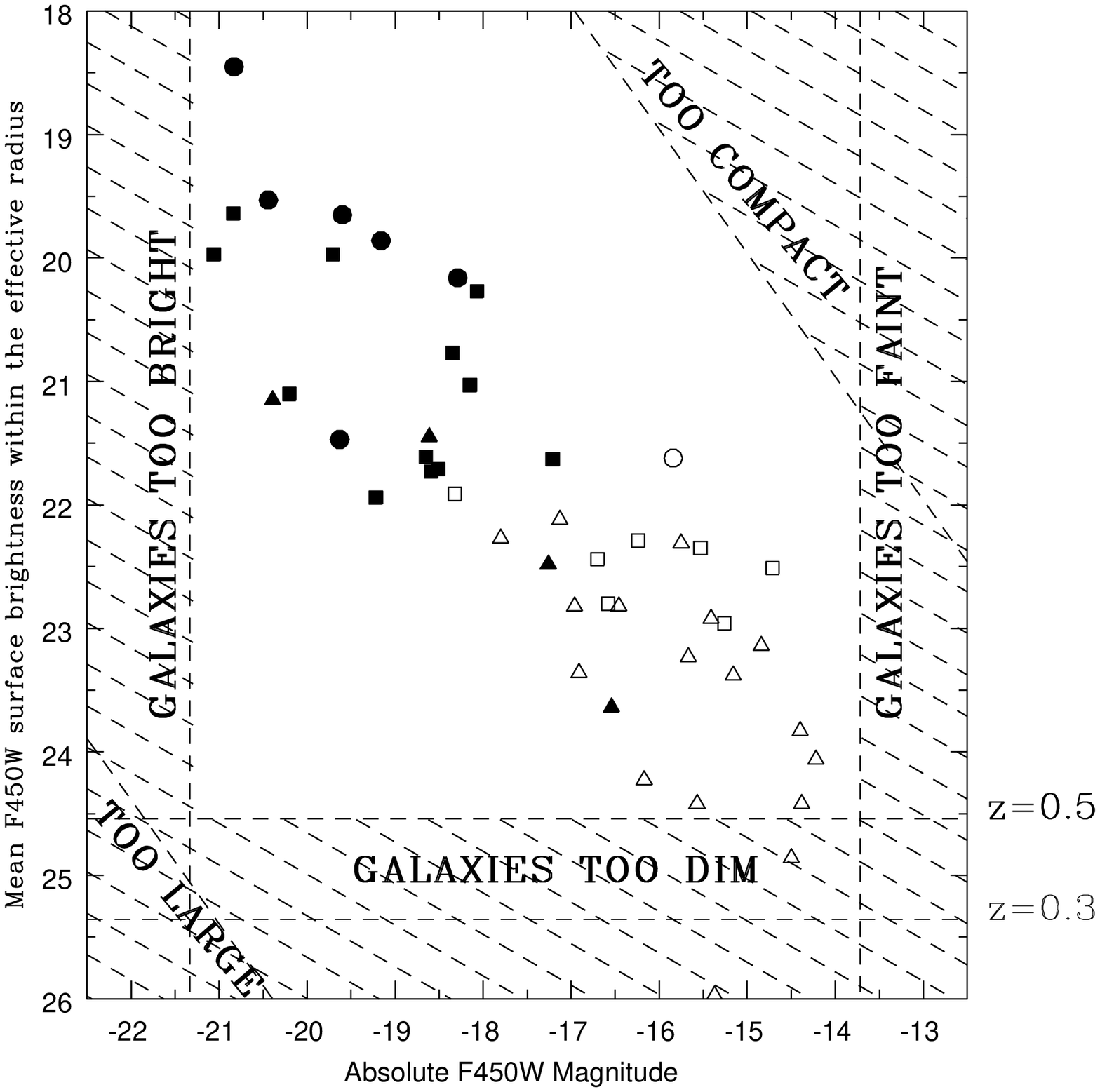,width=16pc}}
\caption{Figures from Driver (1999), showing the utility of the
  bivariate brightness distribution for establishing the bulk
  characteristics of ``local'' ($0.3<z<0.5$) galaxy samples while
  explicitly accounting for selection effects. In both panels open
  circles are early-type galaxies, asterisks are spirals, and
  peculiars are shown as triangles. {\bf [Left:]} Absolute magnitude
  versus photometric redshift for galaxies with $I<27$ mag in the HDF.
  The region enclosed by dashed lines corresponds to the {\em volume-limited} sample
  shown in the next panel. {\bf [Right:]} The bivariate brightness
  distribution for the HDF volume-limited sample defined on the left.
  The lines indicate the selection functions for the HDF survey data.
  Note the clear absence of a dominant population of luminous low
  surface brightness systems, the clear trend for morphology to be
  correlated with both surface brightness and luminosity, and the
  strikingly large area within which systems that are absent could
  have been detected.}
\end{figure}

Including surface brightness as a classification criterion makes
sense on several levels. Firstly, the majority of galaxies in
Universe are low-surface brightness dwarfs whose morphologies lie
off the Hubble sequence. Clearly the the visibility of such
systems is a strong function of the limiting surface brightness of
the observations, and an understanding of the surface brightness
of a given population allows the calculation of reasonably
unbiased selection functions that are necessary to calibrate the
faint-end slope of the galaxian luminosity function. Another
reason why surface brightness might be considered a fundamental
morphological parameter is because of the rather tight
correlations between surface brightness, luminosity, and central
concentration. These allow two (or more) of these parameters to be
used in decision-tree based galaxy classification strategies that
capture much of the variance in the Hubble sequence (Doi,
Fukugita, \& Okamura 1993; Abraham et al. 1994)\footnote{It is
interesting to note that the
  mean surface brightness of early-type dwarf galaxies decreases with
  luminosity, while the mean surface brightness of late-type dwarfs
  increases with luminosity.}. Finally, mean surface brightness
can be viewed as a crude tracer of angular momentum (Heavens
\& Jimenez 1999).

The utility of including surface brightness as a classification
criterion is probably best illustrated in a recent paper by Driver
(1999), the results from which are shown in Figure~3. This work
exploits the extraordinarily deep limiting surface brightness of
the HDF observations to construct a {\em volume-limited} sample of
the nearby Universe that is largely immune to the surface
brightness selection effects whose putative importance has been
the source of so much recent debate (Bothun, Impey, \& McGaugh 1997 and
references therein). Driver concludes that {\em
  luminous} low-surface brightness galaxies are relatively rare
($<10\%$ of the galaxy population), and that dim galaxies at all
luminosities contribute only very modestly to the both the luminosity
density and mass budget of the Universe, in basic agreement with the
results of Ferguson and Babul (1998).

A disadvantage of using surface brightness as a classification
criterion is that its physical interpretation is closely coupled to
assumptions regarding star-formation history (and hence mass-to-light
ratio). However, similar objections can also be raised regarding the
use of asymmetry, and to some extent central concentration, as
fundamental parameters.  In fact the parameter most strongly
correlated with position on the Hubble sequence for local Sa--Scd
spirals is rest-frame colour, and by inference star-formation history.
In the next section I will argue that the connection between
morphology and star-formation history should be made explicit when
probing the high-redshift Universe.

\section{The Morphology--Star-formation Connection}

\begin{figure}
  \centerline{\epsfig{file=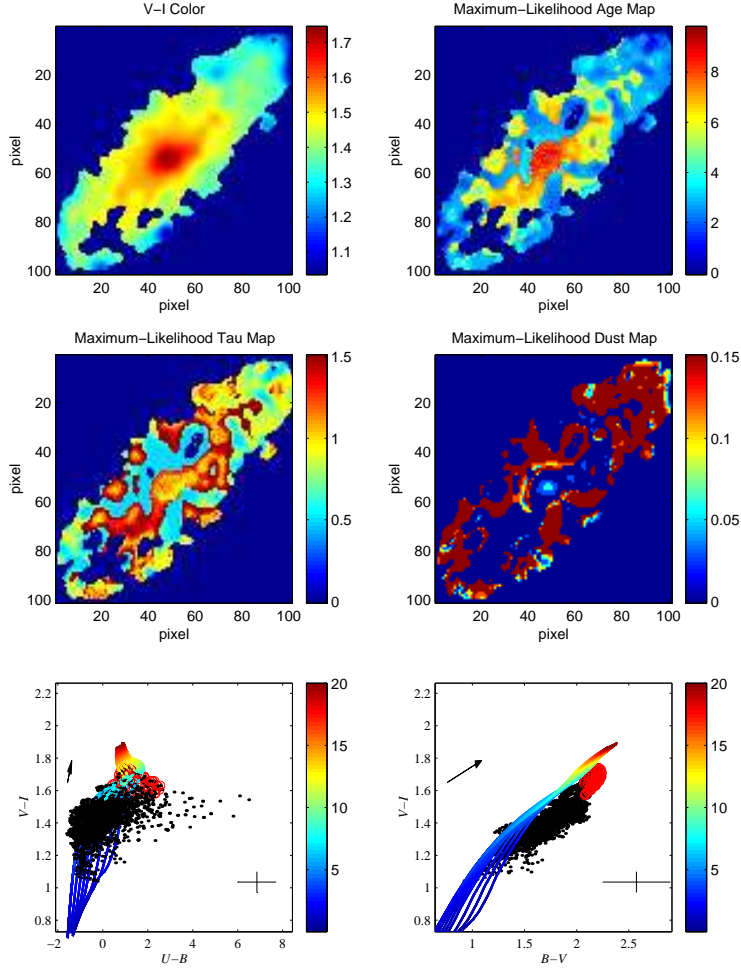,width=25pc}}
  \caption{Internal colour analysis of
    a $z$=0.517 spiral in the HDF.  A $V-I$ colour map for the galaxy
    is shown at top left, and refers to an area 4 $\times$ 4 arcsec
    sampled at the drizzled pixel scale of 0.04 arcsec and limited to
    contiguous pixels contained within the $\mu_B$=26.0 mag
    arcsec$^{-2}$ isophote. The pixel-by-pixels colours in the four
    HDF bands are compared to the predictions of exponential
    star-formation history spectral synthesis models for the redshift
    in question.  The arrows on these plots show a $E_{B-V}=0.1$~mag
    reddening vector calculated from the extinction model of Calzetti
    (1997).  Model tracks for solar metallicity are shown in the
    bottom two panels. Red points refer to pixels within a 5 pixel
    radius of the center of the image. The corresponding age, star
    formation timescale, and dust maps are determined using the
    maximum-likelihood formulation.  Corresponding best-fit maps for
    age, star-formation history e-folding timescale $\tau$, and
    $E_{B-V}$ are shown at top-right, middle-left, and middle-right.
    Figure taken from Abraham et al. (1999a).}
\end{figure}

The notion that morphology is a transient property of galaxies is
a key idea in hierarchical models for galaxy evolution.  This
suggests that the best way forward may be a marriage of stellar
population-based studies (which focus on stellar content that is
preserved during morphological transformations) with Kormendy's
notion of a component-based approach to morphology.  Ideally, such
an approach should avoid the use of both integrated colors (which
fly in the face of stellar population work and ignore the greatest
advantage of HST, namely its ability to resolve distant systems),
and profile fits (because distant galaxies clearly aren't as
smooth and axially symmetric as their local counterparts, and the
locally-defined canonical fitting laws may no longer work).

An example of this approach is shown in Figure~4, which comes from the
pilot study described in Abraham et al. (1999a).  In this paper the
spatially resolved colours of a sample of bright $z<1$ galaxies of
known redshift in the Hubble Deep Field are analyzed by matching
resolved four-band colour data to the predictions of evolutionary
synthesis models (Bruzual \& Charlot 1993). This procedure quantifies
the the relative age, dispersion in age, ongoing star-formation rate,
and star-formation history of distinct components. The central idea
behind this approach is conceptually similar to that used when
applying color--magnitude diagrams to establish the epoch of formation
for galaxy clusters (Bower, Lucey, \& Ellis 1992): {\em dispersion} in
color probes the history of star-formation. By assuming a simple
extinction law, the presence of dust can be tested for and its effects
incorporated into the color modeling. On the basis of this study,
Abraham et al. (1999a) concluded that $\sim 30$\% of early-type
galaxies with $I<23$ mag show evidence of star formation which must
have formed at least 5\% of the stellar mass of the galaxy within the
past third of the galaxian age at the epoch of observation. This
result is largely independent of assumptions with regard to
metallicity and is in agreement with recent spectroscopic observations
of field ellipticals (Schade et al. 1999), and contrasts with the
strikingly low dispersion in the color-magnitude relation of
HST-selected ellipticals in distant rich clusters (Ellis et al. 1997;
Stanford, Eisenhardt, \& Dickinson 1998) .

This methodology can also be used to analyze the relative
histories of bulge and disk stars in spiral galaxies. For example,
the established view is that galactic bulges form at high
redshifts through early dissipationless collapse (Eggen,
Lynden-Bell, \& Sandage 1962; Carney, Latham, \& Laird 1990). This
is based principally on the irrefutable evidence for old stellar
populations concentrated in the bulge of our own Galaxy (Baade
1957; 1963) and contrasts with more recently-developed
hierarchical galaxy formation models (Kauffmann et al 1993, Baugh
et al 1996) where elliptical galaxies form from the merger of
early disk systems which can, in turn, continue to accrete gas to
form a two component spiral galaxy. A safe prediction of the
hierarchical models is that spiral bulges should, on average,
contain older stars than their associated disks which form by
subsequent accretion. Moreover, statistically at a given redshift,
bulges should be older and redder than field ellipticals which
predominantly form from the merger of previously created spirals.
Importantly, these conclusions should remain valid regardless of
the particular cosmological model or initial power spectrum which
governs the rate of assembly of massive galaxies. As such, a
comparison of the relative colors of ellipticals and spiral bulges
offers a remarkably simple, but powerful, test of hierarchical
assembly models (Ellis \& Abraham 1999; Peletier et al. 1999). Of
course a {\em third} alternative for the origin of stellar bulges
proposes their manufacture via various instabilities of
pre-existing disks (see \S2.3, and the recent review by Combes
1999). The point here is that the establishing the relative
importance of the three bulge formation processes as a function of
redshift is clearly another area where morphology and stellar
population studies unite to provide basic tests of galaxy
formation models. Abraham et al. (1999a) concluded that median
ages of bulge stars are  significantly older than those in galactic
discs, and exhibit markedly different star-formation histories.
This is really only the tip of the iceberg --- high-redshift
resolved multi-color studies of this kind should be pushed
into the infrared in order to probe more directly the formation of
stellar mass.

\section{Conclusions and Future Directions}

In this review I have emphasized the transition from the
established morphologies of the local Hubble sequence to the
peculiar forms of the most distant Lyman break galaxies.  While I
have tried to highlight lines of continuity as a function of
redshift, one must be careful not to invent lines of continuity
where they do not exist.  The framework of the Hubble sequence
seems to be a fairly recent phenomenon. It has begun to break down
by $z=0.5$, and by $z=1$ it provides a rather poor general
description of the luminous galaxy population.  At higher
redshifts less emphasis should be placed on classical morphology,
and more emphasis on quantifying the visibility of fundamental
morphological components, and on their resolved stellar
populations.

The  redshift range $0<z<1$ is a particularly interesting for
establishing how the Hubble sequence is built up.  The size
functions for the various classes needs to be more firmly
established, on account of their fundamental role in testing
theory.  The connection between the formation of the Hubble
sequence and the sharp drop-off in the integrated star-formation
history of the Universe needs to be understood. Emphasis also
needs to be placed on developing quantitative methods for the
parameterization spiral structure, perhaps along the lines of the
useful system developed by Elmegreen \& Elmegreen (1982). How does
the density enhancement from spiral structure relate to the
overall star-formation rate? Another valuable contribution would
be the development of a really robust method for distinguishing
between S0 and elliptical systems at high redshifts, in order to
settle conclusively the origin of the Butcher-Oemler effect
(Dressler et al. 1994; Ellis et al. 1997;  Andreon 1998).

At higher redshifts the proportion of morphologically peculiar
systems that are mergers in progress needs to be established. New
metrics, such as global asymmetry and mean surface brightness,
should be combined with dynamical studies in order to investigate
this. Perhaps these metrics could also help answer another
intriguing question: what is the highest redshift at which really
archetypal spiral and elliptical galaxies can be detected?  On the
basis of photometric redshifts and optical HDF data, I can just
about convince myself that there are reasonable looking disk+bulge
systems at $z=1.5$, as well as centrally concentrated early-type
systems at $z=3$. NICMOS observations of the HDF may allow these
issues to be investigated at an appropriate rest wavelength.

Generally speaking, morphological work needs to be pushed further
to the infrared. We need to establish the role of dust, and the
importance of ``old'' stellar populations at high redshift.
Resolved stellar population work in particular really should be
done in the near infrared in order to  probe directly the buildup
of stellar mass without being biased by the presence of relatively
recent star-formation activity dominating the observed rest-frame
ultraviolet flux.  It is interesting that, in many ways, the
motivation for these sorts of programs is similar to the
motivation for multi-colour imaging surveys of {\em local}
galaxies that are now underway. For example, compare the results
presented in \S4 with the scientific rationale for the Ohio State
University Bright Spiral Galaxy Survey (presented at the {\em
last} South Africa morphology meeting) given by Frogel, Quillen
\& Pogge  (1996).  As with so many other areas of astronomy, our
understanding of the distant Universe is limited by our poor
understanding of our own neighborhood.

\acknowledgements I thank David Block and the other organizers of
this meeting for the opportunity to present this review, and for
their indulgence in allowing me to speculate on future directions
for morphological work. Financial support from the sponsors of
this meeting (The Anglo American Chairman's Fund, and SASOL), is
gratefully acknowledged.  I also thank my collaborators on the HDF
and LDSS/CFRS campaigns for many useful discussions, and Simon
Driver for refereeing this review on short notice.

\end{article}

\end{document}